\documentstyle[12pt]{article}
\input{tcilatex}
\input tcilatex

\begin{document}

\begin{center}
{\bf {\large Modelization of the Impedance Spectroscopy of composites by
Electrical Networks}} \bigskip

L. Zekri and N.Zekri\footnote{{\normalsize Regular Associate of the ICTP.}}%
\\[0pt]

{\em U.S.T.O., D'{e}partement de Physique, L.E.P.M.,B.P.1505 El M'Naouar, Oran,
Algeria\\[0pt]}

{\em and\\[0pt]}

{\em The Abdus Salam International Centre for Theoretical Physics, Trieste,
Italy,}\\[0pt]

\bigskip {\em and}\\[0pt]

J.P. Clerc\\[0pt]

{\em Ecole Polytechnique Universitaire de Marseille, Technopole Chateau
Gombert, 13453 Marseille France. } \\[0pt]

MIRAMARE -- TRIESTE\\[0pt]
\end{center}

\vspace{1.5cm}

{\bf Abstract}

\hspace{0.33in} We examined in this work the effect of the inter-grains
distributions as well as the scaling of the complex impedance in order to
analyze its frequency dependance for composite metal-insulator
films. The dependance of the characteristic frequencies on the
inter-grain distribution is shown for large sample sizes. The impedance
spectra become statistically stable above sizes of the order $200$x$200$.

\vspace{2cm}


\newpage

\section{Introduction}

\hspace{0.33in} Composite and granular materials have permitted during last
decades an important enhancement in various materials science fields like
electronics, mechanics and optics.

\hspace{0.33in} Indeed, such systems showed physical properties
unreachable by the conventional materials. Thus some composites
presented actually a very interesting corrosion resistance, which led to
choose them as a skin of boats \cite{Favoto}. Other metal-insulating
mixtures presented an anormally high optical absorption \cite{Gadenne},
thus instigating interest to military industry. We can also mention some
polymer-Carbone black mixtures which reach electrical resistances
comparable to the lead allowing interesting applications mainly on the
fabrication of light batteries \cite{Prigodin}.

\hspace{0.33in} Such performances may depend both on the components
properties, the grains distribution as well as the contact between
them. For instance, Hertz showed that electrical resistance decreases
for an increasing stress. From the various techniques used for analysing
the physical properties of such materials we can mention the impedance
spectroscopy. This method allows to determine the frequency
dependance of the electrical response and particularly the
characteristic frequencies. These frequencies inform us about the
dielectric and optical properties of these materials.

\hspace{0.33in} Recently, a numerical and experimental investigation of
impedance spectra on powder \cite{Tedenac} determined their
characteristic frequencies, by taking into account tunnelling effect
between grains. In this study, the conductance was taken to depend
exponentially on the inter-grains distance whose distribution was chosen
to be uniform.

\hspace{0.33in} In this work, we use the above modelization by
assuming the inter-grain distribution to be Gaussian in order to
examine the effect of the conducting grains organization. We
investigate also the scaling effect on the complex impedance at the
characteristic frequencies.

\section{The model}

\hspace{0.33in} Impedance measurements on mixtures of conducting and
insulating powder showed that the system can have an inductive behavior.
This behavior is attributted to the inter-grains contact rather than the
grains themselves in this range of frequencies. The system is then modelled by $RLC$ 
circuits representing the contact impedance \cite{Zeng, Clerc}. Some polymers doped
by carbon black have shown an electric conduction even with an
insufficient carbon concentration to make an infinite cluster. The
resulting current is either a quantum (tunnelling) effect or a thermal
activation energy. Because of the variation of the inter-grains distances,
the conducting or insulating bonds do not have identical conductances.

\hspace{0.33in} Let us consider two agregates conductors bonded by $N(p)$
tunnel junctions, $p$ being the concentration of metallic grains. On a
network of bonds $N(p)$ is the number of simple dielectric junctions (bonds)
allowing eventually a tunnelling transmission between two nearest neighbor
agregates. The conductivity of the $i^{th}$ tunnel junction then reads \cite
{Abeles, Sheng}

\begin{equation}
\Sigma_{i} \simeq a_{0}\sigma_{m}\exp (-l_{i}/\delta)
\end{equation}

\noindent where $a_{0}$ is a typical grain (or agregate) size, $l_{i}$ the
size of the $i^{th}$ junction ($0<l<a0$) and $\delta$ the
characteristic size for tunnelling effect. This form can be obtained by
analogy with the electron transmission by thermal activation above a 
barrier energy $E$ ($exp(-E/KT)$, $K$ being the Boltzman constant and
$T$ the temperature). Equation (1) becomes then

\begin{equation}
\Sigma_{i}\simeq a_{0}\sigma _{m}\exp (-\lambda _{0}x_{i})
\end{equation}

\noindent with $\lambda _{0}=a_{0}/\delta$ and $x_i=l_i/a0$ is a
dimensionless distance distributed between $0$ and $1$. Only the best
junctions contribute to the tunnelling current. We can then assume that
the tunnel conductance between two conducting agregates $j,k$ is 

\begin{equation}
\Sigma_{j,k}\simeq a_0\sigma _m\exp (-\lambda _0x_{\min \{j,k\}}).
\end{equation}

\noindent When the percolation $p_c$ \cite{Stauffer} is reached, the
distance between two agregates becomes more and more smaller and the
probability of finding a junction of high conductance increases. By
taking into account all junctions the conductance becomes

\begin{equation}
\Sigma_{j,k}\simeq a_0\sigma _m\exp (-\lambda
(p)y_{j,k})\,;\,0<y_{j,k}<1
\end{equation}

\noindent with $\lambda (p)\simeq \lambda_0/N(p)$ a $p$ dependent
renormalized tunnelling parameter. If we simplify the fluctuations on $C$
(considered constant here), the quantities $R$ and $L$ obey to the
following distributions (From Eq. 4)

\begin{eqnarray}
R_{i,j} &=&R_010^{\lambda x_{i,j}} \\
L_{i,j} &=&L_010^{\lambda x_{i,j}}
\end{eqnarray}

\noindent with $x_{i,j}$ a random variable distributed within the intervalle 
$\{{0,1\}}$, $R_0$ is the minimum value of $R$ and characterises the doping
of the system and $\lambda $ the width of the distribution. From
experimental data, there is a constant ratio between $R_{i,j}$ and
$L_{i,j}$. The computation of the effective conductance is done by using
Kirchoff equations following a method solving exactly these equations
\cite{Zekri}. For a system of size $L$x$L$ this leads to the use of an
$L^2$x$L^2$ matrix which is handled by blocs because of its sparse shape and
arrangement near the diagonal region. This method allows us to reach 
very large sizes in comparison with those reached by \cite{Tedenac} and to
study the scaling effects of the impedance.

\section{Results and discussions}

\hspace{0.33in} Impedance spectra measurements on powder can be  realized
via an impedance analyzer where the sample is constituted actually 
of insulating grains (like galss balls treated at surface) and conducting ones
(like glass balls leads treated at surface) \cite {Tedenac}. 

\hspace{0.33in} In a theoretical work \cite{Tedenac}, the inter-grains
distance was uniformly distributed assuming that there are no
correlations between them. However, because of the boundary
conditions on such systems (the size of the system is fixed) it seems
reasonable to think about a distribution taking into account correlated
distances which are characterized by their average value with
fluctuations around it. We choose for this end a Gaussian distribution
with the same parameters: $R_{0}$, $C$, $L_{0}$ used by ref.
\cite{Tedenac} and $\lambda =7$.

\hspace{0.33in} Figure 1 shows a comparison between these distributions
on the complex impedance spectra for a sample of size $30$x$30$. We define
here the characteristic frequencies as those corresponding to the maximum
of the real part of the impedance ($\omega_r$) and its imaginary part
($\omega_i$). In a simple $RC$ branch $\omega_{i}$ is the relaxation
frequency while in for an $RLC$ circuit $\omega_{r}$ is the resonance 
frequency. These frequencies seem to depend slightly
on the inter-grains distribution. This implies a slight effect of the
inter-grains distributions on the optical and dielectric properties of
the system. However, these results correspond only to small sizes 
($30$x$30$) which are probably statistically insufficient.

\hspace{0.33in} It is then important to investigate the scaling effect of
these spectra. We remark in figure 2, large fluctuations for small sizes
which decrease for larger samples. Except for the above fluctuations, it
is difficult to study the scaling effect from the whole range of
frequencies.

\hspace{0.33in} We examine then the scaling effect only on the
characteristic frequencies which represent the main physical properties
of the system. This is shown in figure 3 where, despite of the
fluctuations for small sizes there is an obvious scaling dependence for
both parts of the impedance with a saturation at a size $200$x$200$ for
$\omega_{r}$ while for $\omega_{i}$ it should be reached at a larger
size (around $400$x$400$). From this figure it is clearly seen that
$\omega_{r}$ depends on the inter-grains distribution (where the uniform
distribution yields a larger frequency) while $\omega_{i}$ seems to be
independent (within the statistical errors).

\section{Conclusion}

\hspace{0.33in} We have studied numerically the impedance spectra for
metal-insulator composites in order to model their dielectric properties.
The inter-grains distances distribution seems to affect sensitively one 
characteristic frequency ($\omega_{r}$) for large samples while it does 
not seem to change the other frequency ($\omega_{i}$). The scaling investigation  
showed that $\omega_{r}$ saturates above sizes of $200$x$200$ while we need much
larger samples (about $400$x$400$) to reach the saturation of the other
frequency. It is then interesting to search the distribution which fits best the 
the existing experimental data particularly the characteristic frequency 
since from the comparison between the two used 
distributions the spectrum seems to be not globally affected.  It is
also important to investigate the spectra of such systems away from the
percolation threshold in order to apply these results to the corrosion and
also to the control of active systems composition (insulator-insulator) for
pharmaceutical applications \cite{Macdonald, Duncan}.

\noindent{\bf Acknowledgement}

\hspace{0.33in} Two of us (L.Z. and N.Z.) would like to acknowledge the hospitality 
of the Abdus Salam ICTP (Trieste, Italy) during the progress of this work. Financial 
support from the Arab Fund is acknoledged by N.Z.

\newpage

\begin{center}
{\bf Figure Captions}
\end{center}

\bigskip

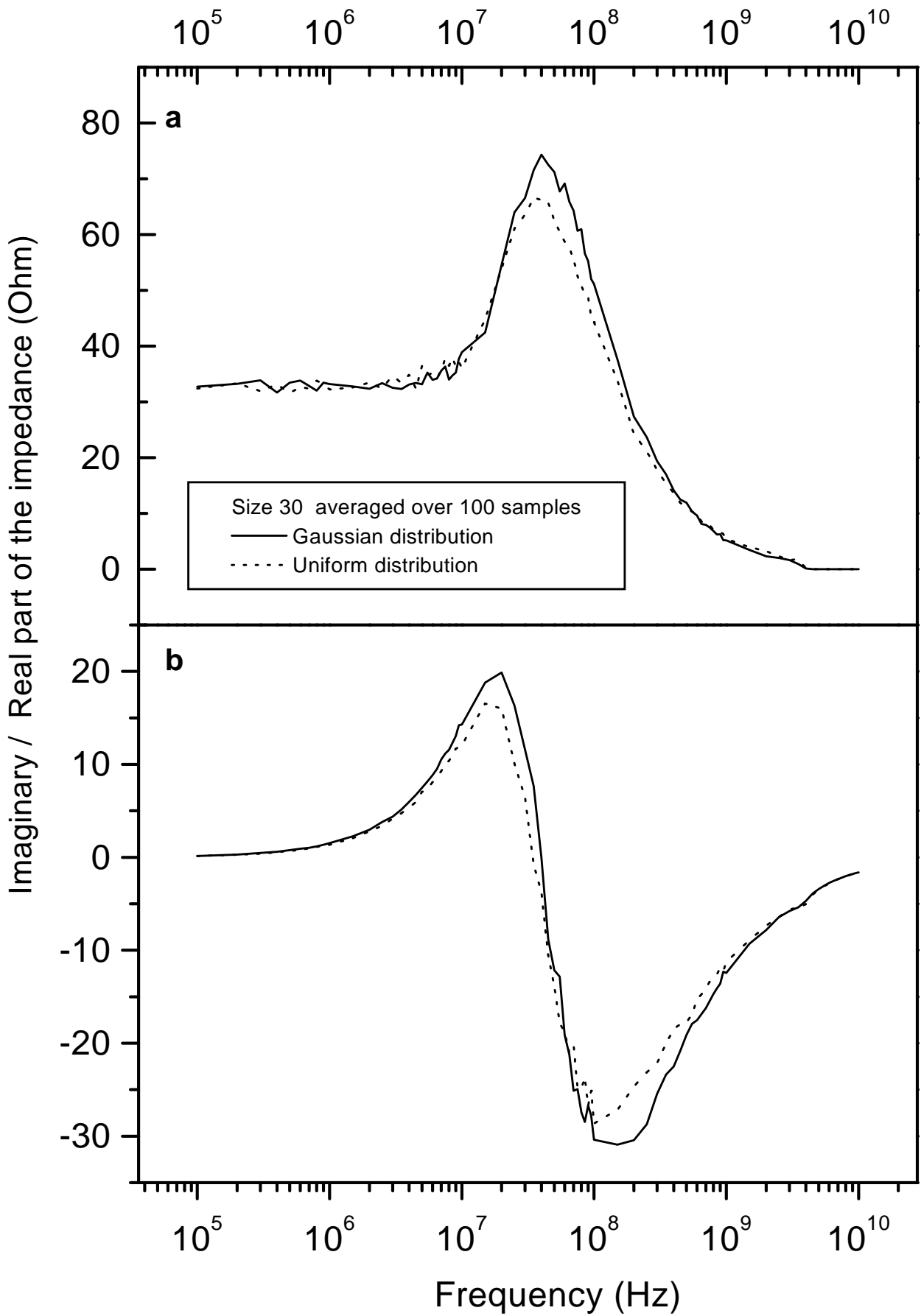
{\bf Figure 1} Comparison of the real part (a) and imaginary part (b)
spectra of the impedance for two different distributions.

\bigskip

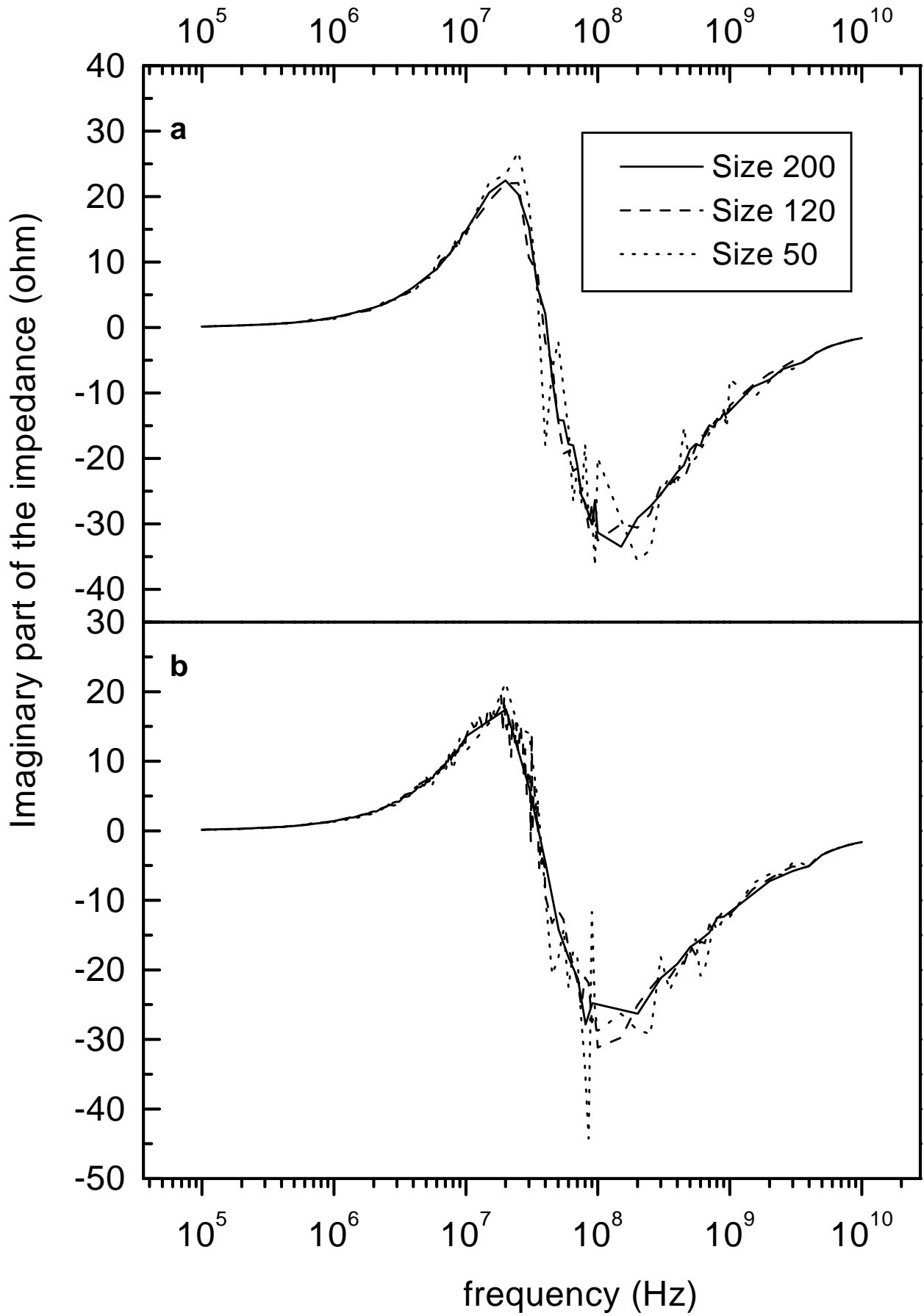
{\bf Figure 2} Imaginary part of the impedance for different sizes and for a
uniform distribution (a) and a Gaussian distribution (b).

\bigskip

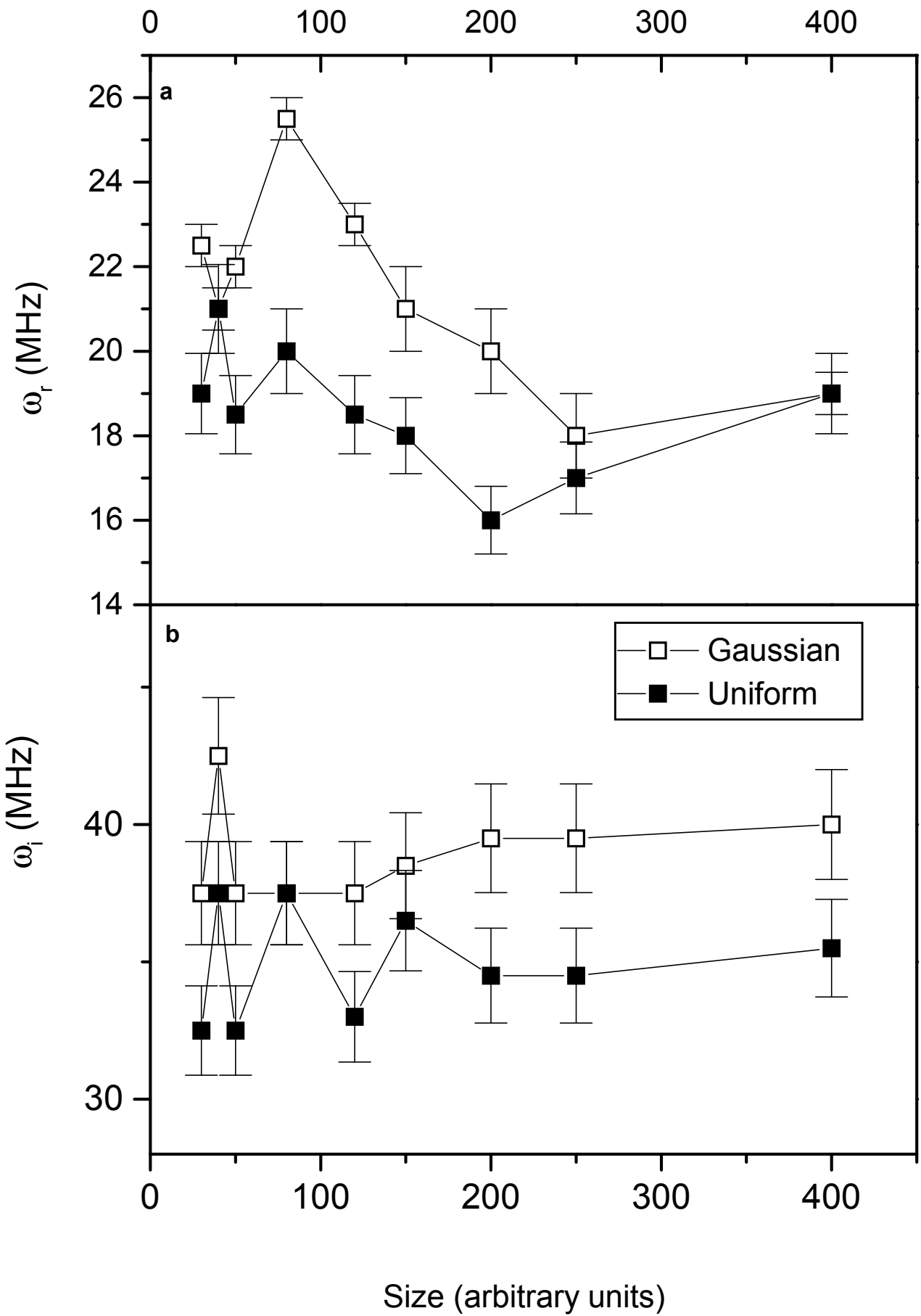
{\bf Figure 3}Comparison of the characteristic frequencies $\omega _i$ (a)
and $\omega _r$ (b) for uniform and Gaussian distributions. 

\end{document}